Political Influence and Corporate Profits: A Study of Hungarian Firms

Zoltán Bartha, zoli@ekon.me

Abstract

This paper investigates the extent of political rent seeking in Hungary in the 2010s. Political capitalism —where powerful private interests influence public policy for private gain—creates opportunities for rent seeking that vary across sectors. The analysis is based on a theoretical model assuming rent seeking occurs in a three-stage process: changes in economic institutions granting regulatory privileges, which are enhanced by political-business networks; this leads to scarcities, and increased market power in certain markets; which then generates rents. To quantify this, the study evaluates Hungarian political capitalism by examining the impact of political decisions on firms' rents, analysing the profit trends of the 1,000 largest Hungarian firms (selected annually by net sales) and comparing their mean profit share (earnings before tax) across two periods: 2008–2012 and 2019–2023. A significant increase in a sector's mean profit share was assumed to indicate increased rent seeking. Using Welch's two-sample t-tests, three sectors were identified as potentially experiencing increased rent seeking: agriculture, construction, and financial and insurance activities. Quantitative findings include a 320% increase in mean agricultural profit share (70% in mean ROA), a more than fivefold increase in construction mean profit share (mean ROA from 3.3% to 10.1%), and a more than 6.5 times increase in financial sector mean profit share. Furthermore, a similar Czech analysis showed no significant increases in any sector's profit share, suggesting that the detected rises in Hungarian sectors are linked to domestic activities rather than external factors, which strengthens the findings.

1. Introduction

Capitalist systems fundamentally rely on private property rights, including the right to trade assets, for the allocation of aggregate wealth. The primary distribution of income in such systems is mostly determined by decisions made in markets. Economic theory often presents an idealized model of market capitalism where income distribution is determined almost purely by competitive market transactions, a scenario widely associated with the efficient allocation of resources (Arrow & Debreu, 1954) and conducive to economic development. Yet, this textbook ideal contrasts with the reality of many capitalist economies where primary income distribution is also significantly impacted by non-market decisions rooted in political processes. Systems where this political influence on income distribution is pronounced are often termed political capitalism. Political capitalism can distort resource allocation and potentially undermine long-term economic growth.

Political capitalism usually emerges from a complex symbiosis between state and non-state actors, a dynamic that can manifest in various ways. On one hand, it may arise from powerful private agents capturing weak state institutions (Ganev, 2009). Conversely, political capitalism can also be created by powerful political elites (Milanovic, 2020). According to Milanovic (2020, p. 17), this latter form is characterized by three key features: 1) the state is managed by a technocratic bureaucracy primarily focused on delivering economic growth and rising incomes; 2) laws and rules are applied selectively or arbitrarily to benefit the ruling elites; and 3) the state is free of legal constraints. Regardless of its specific origin, in political capitalism, political decisions often override market outcomes. This occurs either because decisions directly serve the interests of political elites or because business elites successfully leverage their influence over the state. One common strategy resulting from such political influence is the deliberate creation of scarcities in markets, enabling those controlling scarce assets to command a premium and thus generate rents. The pursuit of these unearned gains, an activity that benefits individuals or specific groups but is detrimental to society overall by distorting resource allocation, is termed rent seeking (Laband & Sophocleus, 2019).

This paper examines the state of Hungarian political capitalism through the lens of rent seeking. Rent seeking is investigated using an indirect approach: analysing changes in firm profitability. The analysis focuses on the top 1,000 Hungarian firms based on net sales. For each firm, its share of the total profits generated by these 1,000 firms is calculated across two distinct periods. The first period spans 2008–2012, and the second, more recent, period covers 2019–2023, utilizing the latest available data. The study then identifies the sectors in which firms experienced a rise in their collective profit share among the top 1,000. This increase is interpreted as potentially indicating the capture of extra rents facilitated by political decisions. In order to ensure that the increase in profit shares is not merely a result of unexamined external factors, this paper compares the Hungarian profit data with corresponding Czech data.

This study makes two main contributions to the literature on Hungarian political capitalism. First, it employs a method for measuring rent seeking that has not been used in previous studies. Prior Hungarian empirical attempts have primarily focused on contracts won through the public procurement system. Second, it identifies specific sectors where rent seeking may be suspected based on profitability data. These sectors can then be compared with those already investigated by other researchers in this field.

The remainder of this paper is organised into six sections. Following this introduction, the next section provides an overview of the rent seeking literature, focusing on the institutional environment of the process and the efforts taken to measure it. Section 3 introduces the theoretical model, and Section 4 presents the data used for the analysis. Sections 5 and 6 introduce and discuss the results. Finally, Section 7 concludes the study with summarising remarks.

2. Literature review

To assess the impact of political capitalism on rent seeking in Hungary, two topics need to be presented. First, rent seeking needs to be clearly defined and its connection with political capitalism explained. Second, the methods of measuring rent seeking have to be discussed. The literature review is therefore divided into two subsections.

2.1. Identifying rent seeking in political capitalism

The concept of rent seeking was introduced to the economics literature by Gordon Tullock in his 1967 paper, where he discussed "the amount of effort that will be invested in attempting to obtain a monopoly" (Tullock, 1967, p. 225). The term itself was later coined by Anne O. Krueger. In her 1974 paper, Krueger described rent seeking as an economic activity where firms devote resources to competing for government-generated rents (Krueger, 1974, p. 293). Activities focused on rent seeking do not generate new value; as Jagdish N. Bhagwati termed them, they are "directly unproductive, profit-seeking" activities (Bhagwati, 1982). Rent seeking is considered socially wasteful because it reallocates resources away from productive uses towards unproductive ones (Sobel & Garrett, 2002). The social costs of this activity include the opportunity cost of the resources invested (Del Rosal, 2011).

In an economy where goods and resources are primarily allocated by markets, the extent of rent seeking is determined by economic institutions, which are themselves shaped by political decision-makers (Acemoglu & Robinson, 2012). Rent seeking occurs when political decision-makers are incentivised to alter economic institutions in ways that reduce market competition (Aidt, 2016). Less competition increases scarcity and subsequently drives up prices, thereby generating extra revenue and profits for firms without any corresponding increase in social wealth. In fact, rent seeking has a negative macroeconomic impact by hurting the income of both workers and capital owners (Christou et al., 2025). The motivations of political decision-makers can vary, but in a system of political capitalism they are primarily related to the vested interests of the political elites.

Within an institutional framework, rent seeking can be interpreted as a three-stage process:

A) A change in economic institutions.
B) This ultimately leads to increased scarcities.
C) These scarcities, in turn, generate rents.

Stage A is generally associated with two inter-related phenomena: A1) Regulatory privileges; and A2) Political-business networks. Regulatory privileges (A1) are often referred to as market capture (Mihályi & Szelényi, 2019; Zywicki, 2015), or the manipulation of public policy (Chen et al., 2011). Various types of regulatory privileges are discussed in the literature, including exclusive licenses, subsidies, favourable regulations (Aidt, 2016), and tax credits (Chen et al., 2011). Assetisation and enclosure is a special case of regulatory privileges. It involves regulation that permits firms to convert resources or rights into income-generating assets. These assets then produce income through exclusive control (Birch & Ward, 2024; Sayer, 2023). One example of this is the series of policies implemented in Hungary to regulate land ownership, which led to a concentrated owner structure. This structure subsequently allowed a small group of wealthy landowners to earn rents from European Union land-based subsidies (Gonda, 2019).

These privileges are all entirely compatible with political capitalism. In addition to these, there are tools specifically common within the Hungarian model: the transfer of ownership to preferred groups via temporary nationalisation, and state credits and selective favours achieved through the regular use of special rights, such as declaring investments of national economic relevance (Csaba, 2022a).

Political-business networks (A2), or clientelism (Cvejic, 2016), both reinforce and normalise rent seeking at an institutional level (Schoenman, 2014). The Hungarian political capitalism is characterised by economic patriotism (Gerőcs & Szanyi, 2019), which favours selected national champions. In this system, politically loyal capitalists effectively subsume domestic businesses (Rogers, 2024). Scheiring (2021) refers to the Hungarian system as the accumulative state, where stability is sustained by an alliance between the political elite and national capitalists, alongside multinational capital–an influential interest group also highlighted by Orenstein & Bugarič (2022). In a system of political capitalism, rent dissipation may be even lower due to institutional barriers to entry in rent seeking. Holcombe points out that policymakers—the most influential group in political capitalism—benefit most when they limit competition for rents. This results in a large surplus to be divided between the creators and the recipients of these rents (Holcombe, 2017).

Regulatory privileges create artificial barriers to entry (Aidt, 2016), limit competition (Zywicki, 2015), and contribute to the entrenchment of non-competitive practices (Schoenman, 2014). This limited competition enhances scarcity (B), which in turn allows firms to raise prices. The resulting higher prices generate greater revenues for the favoured firms, which is how rents are achieved (C). Limited competition makes it more difficult for consumers to make meaningful comparisons regarding quality and prices, which also favours rent seekers (Di Liddo & Giuranno, 2024). Since the costs of rent seeking are not entirely dissipated (see next section), these higher revenues lead to higher profits for the affected firms.

## 2.2. Measuring rent seeking

Theoretically, rent seeking might appear straightforward to measure using the concept of the dissipated rent, often visualized as the Tullock rectangle representing the potential monopoly profit (Tullock, 1967). However, Tullock himself later admitted that finding measurable costs associated with this potential rent proved extremely difficult in practice (Tullock, 1997). The potential rent represented by this rectangle can be extremely large in real-world situations, yet empirical evidence suggests that firms' investments in rent-seeking activities fall far short of this amount. This discrepancy constitutes the Tullock paradox. Tullock suggested that a significant portion of the potential rent may not be fully dissipated because voters or consumers who ultimately bear the costs may be unaware due to poor, and asymmetric information (Tullock, 1997). Hillman & Katz (1984) demonstrated that if rent seekers are risk averse, or if there are barriers to entry in contests, full dissipation of rents does not occur. Ursprung (1990) showed that incomplete dissipation can be associated with the public good nature of contested rents. Dougan and Snyder (1993) concluded that redistributive policies can be designed in ways that minimise the wasteful dissipation of transfers.

Rent seeking can manifest in several main forms, each potentially requiring different measurement approaches. One primary form involves direct payments to decision-makers (e.g., bribes). A second form includes in-kind compensation, such as expensive dinners, prestigious club memberships, and other similar benefits (Mixon et al., 1994). A third category is indirect rent seeking, where lobbyists attempt to influence decision-makers by shaping public opinion through policy papers and the media (Sobel & Garrett, 2002). Other forms also exist, including the hiring of politicians' friends and family, or investing in political careers with the aim of directly controlling rent-generating power (Laband & Sophocleus, 2019).

Beyond these individual methods, Congleton (2024) highlights that rent seeking contests are systemic and interconnected. Using the democratic system as a model, he illustrates how a single policy innovation can trigger four linked contests: external efforts to influence new regulations, internal competition for pivotal government positions, and subsequent budgetary battles over the tax revenues generated by these very policies. These linkages ultimately scale the prizes available, as competitive efforts both inside and outside government shift to capture the resulting economic rents.

Del Rosal (2011) identified three main methods for measuring rent seeking. The first group comprises indirect measures, which are based on attempts to quantify rents. An example is calculating the rent of protectionism by comparing domestic and world market prices (Krueger, 1974). This method is only valid if full dissipation of rents is assumed, which, based on the previously discussed Tullock paradox, is unlikely to hold in reality. If full dissipation was true, investigating corporate profits would be unproductive, as rents would not be reflected in reported profits.

The second group is referred to by Del Rosal as the accounting perspective. Studies adopting this method assume that certain actions are undertaken for rent-seeking purposes, and thus expenses incurred for such purposes are considered measures of rent seeking. One influential work by Sobel and Garrett (2002), for instance, compared U.S. state capitals with non-capital cities and examined sectors that were significantly more dominant in capitals. They assumed that this dominance was attributable to rent seeking. Another relevant study by UNCTAD investigated the surplus profit of publicly listed non-financial firms across 56 developed, developing, and transition economies between 1995 and 2015. They measured the gap between actual and benchmark (or typical) profits and found significant rents in many sectors (UNCTAD, 2018). The approach I utilise in this paper aligns with this second methodological group.

The third and final group focuses on aggregate measures. It assumes that because rent seeking reallocates resources to unproductive purposes, this slows the pace of economic development. A common approach within this group uses the number of law firms or law college enrolments as a proxy for rent seeking intensity. Regression analysis is then typically employed to examine its relationship with the rate of economic growth (Ebeke et al., 2015; Laband & Sophocleus, 1988; Murphy et al., 1991).

Studies focusing on rent seeking in Hungary have adopted diverse approaches. Szanyi (2022) provides a comprehensive description of the Hungarian economy primarily relying on political economy and institutional concepts. He identifies several key areas of rent seeking: public procurement contracts (particularly substantial in the construction and communication sectors) that favour firms loyal to the ruling elite, renationalization of firms (especially in the utilities and financial sectors), and the capture of the energy, telecommunication, and media sectors.

Rent seeking in the banking sector has also been a subject of study. Nationalization efforts in this sector, and the use of regulatory tools to facilitate them, are discussed by Oellerich (2022), and Toplišek (2020). Toplišek (2020) notes, for example, that foreign ownership in the Hungarian banking sector decreased from 80% to 50% by the end of 2017. Toplišek's work also examines the nationalization of the energy sector in Hungary.

The agricultural sector is a frequent focus for studies investigating rent seeking in Europe (Czyżewski & Matuszczak, 2018; Fałkowski & Olper, 2014; Furtom et al., 2009). Research specifically on Hungary indicates that over 80% of CAP subsidies constituted political rents, which is not unusually high within the EU; these rents primarily accrued to large and very large farms (Czyżewski & Matuszczak, 2018).

Empirical analyses of public procurement provide further evidence of rent seeking. Tóth and Hajdú (2018) analyse over 100,000 public procurement procedures conducted between 2010 and 2016. Their findings document clear favouritism, observing significantly lower intensity of competition for tenders awarded to firms owned by some of the closest allies of the Hungarian prime minister.

3. Theoretical model

The theoretical framework for this analysis is presented in Figure 1. This model builds upon the arguments detailed in Section 2.1. Rent seeking can be conceptualised as a three-stage process. Stage A involves a change in economic institutions enacted by political decision-makers. These changes may grant a variety of regulatory privileges (A1), such as exclusive licences, subsidies, favourable regulation, tax credits, state credit, exclusive control over assets, government intervention into ownership structure, the use of special rights etc. These privileges are further reinforced by the institutionalised norms of political-business networks (A2). Both A1 and A2 are notably prevalent within the Hungarian system of political capitalism (Csaba, 2022b; Szanyi, 2022).

The changes in economic institutions are highly selective, resulting in a strong sectoral bias in their impact. Owing to the uneven distribution of privileges granted by political decision-makers, certain sectors become more susceptible to rent extraction. This leads to the artificial creation of scarcities in some markets more than others (B). This sectoral bias is well-supported by both international studies (Angelopoulos et al., 2021; Hellman et al., 2000) and literature focusing on Hungary (Csizmady & Kőszeghy, 2022; Oellerich, 2022; Szanyi, 2022).

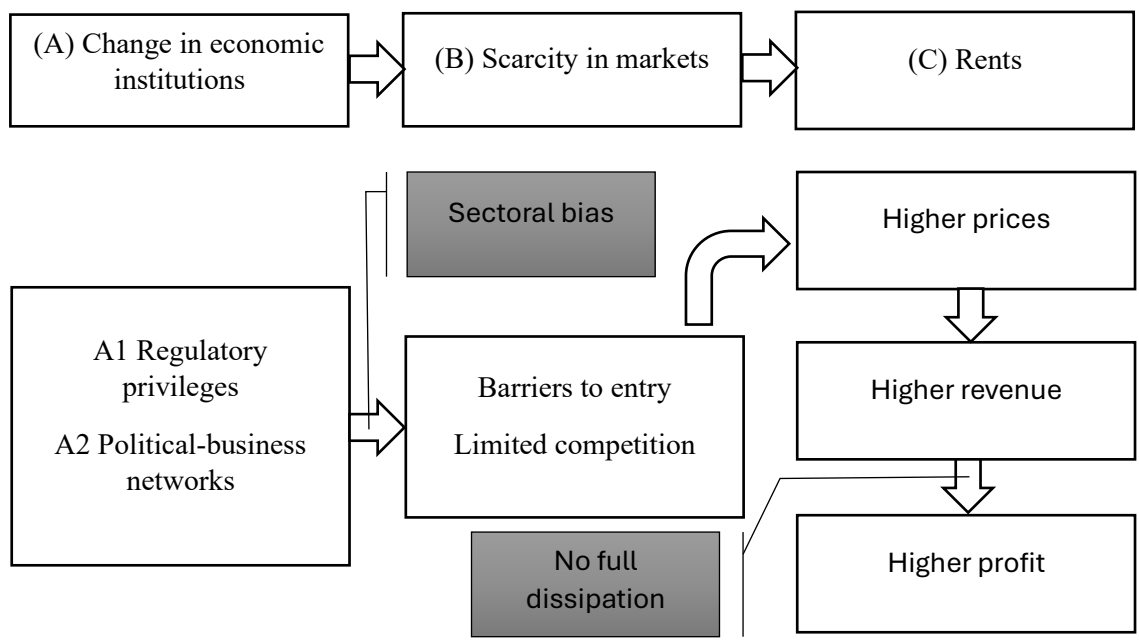

Figure 1. Theoretical model of identifying rent seeking through corporate profits

Source: own work

Firms operating in markets with artificially created scarcities are able to command higher prices, resulting in greater revenue and, consequently, higher profits (C). Higher profits may also arise from macrotrends (e.g., geopolitical, supply-side, or demand-side shifts, or changes in the market power of large corporations), or from firm-level initiatives (e.g., marketing or innovation efforts) that yield competitive quasi-rents. This study addresses these two latter sources of higher profits through a two-pronged control strategy: the impact of general trends is accounted for by comparing Hungarian results with Czech ones, and firm-level developments are controlled for by aggregating profits at a sectoral level.

The primary objective of this study is to detect the distortive effect of A1 and A2 in Hungary through the analysis of sectoral-level profit shifts. This approach is expected to yield convincing results, provided three conditions hold:

1. A1 and A2 are so systematically and powerfully embedded within Hungarian political capitalism that they exert a clear distortive impact on markets.
2. The institutional changes via A1 and A2 have a significant sectoral bias, thereby creating substantially more scarcities (B) in specific markets.
3. Companies belonging to sectors prone to rent seeking do not pay the full costs of the rents they enjoy (i.e., there is no full dissipation of rents).

These three conditions can be considered realistic, as they are supported by the existing literature.

4. Data and methods

My analysis is based on publicly available financial data taken from the CrefoPort database. For each analysed year, the 1,000 firms with the highest net sales were selected. These top 1,000 firms represent a significant segment of the Hungarian economy: in 2023, they employed 21% of the workforce, and their aggregate net sales were 1.5 times the national GDP. Firms were classified into sectors according to the NACE codes of their principal activity. In 2 to 3 percent of cases, NACE codes were missing, requiring manual input. The sectoral distribution of the top 1,000 firms is unbalanced; most companies belong either to the manufacturing or the wholesale and retail sectors (see Table 1). The composition of the top 1,000 firms naturally evolves over time, yet there is substantial overlap between the lists from different years. Specifically, 445 companies from the 2008 top 1,000 list also appear on the 2023 list.

Table 1. Descriptive data of the top 1,000 Hungarian firms

| Year | 2008 | 2009 | 2010 | 2011 | 2012 | 2019 | 2020 | 2021 | 2022 | 2023 |
|---|---|---|---|---|---|---|---|---|---|---|
| Persons employed (thousands) | 703 | 700 | 713 | 728 | 727 | 762 | 806 | 812 | 821 | 782 |
| Total net sales (trillion HUF) | 48 | 44 | 44 | 47 | 47 | 80 | 86 | 96 | 140 | 110 |
| Total earnings before tax (billion HUF) | 1 616 | 1 903 | 1 716 | 1 674 | 1 672 | 3 089 | 2 876 | 5 695 | 6 367 | 6 611 |
| Hungarian GDP (trillion HUF) | 27 | 26 | 27 | 28 | 29 | 48 | 49 | 56 | 66 | 76 |
| **Distribution of companies according to principal sector of activity (percent of total)** | | | | | | | | | | |
| A - Agriculture, Forestry and Fishing | 1.1 | 0.9 | 1.1 | 1.2 | 1.2 | 0.9 | 0.8 | 0.8 | 0.9 | 0.9 |
| B - Mining and Quarrying | 0.6 | 0.7 | 0.6 | 0.6 | 0.6 | 0.4 | 0.4 | 0.3 | 0.3 | 0.3 |
| C - Manufacturing | 32.4 | 31.3 | 32.7 | 34.8 | 34.6 | 34.8 | 34.7 | 36.3 | 37.1 | 37.8 |
| D - Electricity, Gas, Steam and Air Conditioning Supply | 7.0 | 7.5 | 7.1 | 5.9 | 5.9 | 4.3 | 4.7 | 5.0 | 5.1 | 4.9 |
| E - Water Supply; Sewerage, Waste Management and Remediation Activities | 1.0 | 1.1 | 1.2 | 1.5 | 1.5 | 0.8 | 0.8 | 0.9 | 0.7 | 0.7 |
| F - Construction | 4.8 | 5.6 | 5.6 | 4.2 | 4.3 | 6.7 | 7.0 | 6.2 | 5.8 | 6.9 |
| G - Wholesale and Retail Trade; Repair of Motor Vehicles and Motorcycles | 28.6 | 28.5 | 28.1 | 29.5 | 29.6 | 33.0 | 32.0 | 32.3 | 30.6 | 28.9 |
| H - Transportation and Storage | 5.5 | 5.3 | 5.5 | 5.3 | 5.3 | 4.5 | 5.0 | 5.4 | 6.2 | 5.7 |
| I - Accommodation and Food Service Activities | 0.6 | 0.6 | 0.8 | 0.6 | 0.6 | 0.7 | 0.3 | 0.4 | 0.6 | 0.9 |
| J - Information, Communication | 3.2 | 3.7 | 3.8 | 3.6 | 3.6 | 3.8 | 4.3 | 4.2 | 3.7 | 4.1 |
| L - Financial and Insurance Activities | 4.9 | 5.1 | 4.4 | 3.8 | 3.8 | 2.8 | 2.4 | 2.3 | 2.8 | 1.8 |
| M - Real Estate Activities | 3.2 | 2.3 | 1.8 | 2.1 | 2.1 | 0.3 | 0.8 | 0.2 | 0.2 | 0.7 |
| N - Professional, Scientific and Technical Activities | 3.7 | 3.7 | 3.6 | 3.3 | 3.3 | 2.8 | 3.3 | 2.8 | 2.9 | 2.5 |
| O - Administrative and Support Service Activities | 3.0 | 3.1 | 2.9 | 2.9 | 2.9 | 3.5 | 3.1 | 2.5 | 2.7 | 3.5 |
| R - Human Health and Social Work Activities | 0.2 | 0.4 | 0.6 | 0.5 | 0.5 | 0.1 | 0.1 | 0.2 | 0.0 | 0.1 |
| S - Arts, Sports and Recreation | 0.2 | 0.2 | 0.2 | 0.2 | 0.2 | 0.4 | 0.2 | 0.2 | 0.3 | 0.2 |
| T - Other Service Activities | 0.0 | 0.0 | 0.0 | 0.0 | 0.0 | 0.2 | 0.1 | 0.0 | 0.0 | 0.1 |

Source: own calculations based on CrefoPort data; Hungarian Central Statistics Office (GDP)

For each selected year, the share of each company from the total earnings before tax (EBT) of the top 1,000 firms was calculated. The research question guiding this study is whether there was a significant increase in the share of profits (EBT) for companies within a certain sector between the 2008–2012 and 2019–2023 periods. Detecting a significant increase could indicate the impact of political decisions in that sector, suggesting a rise in rent seeking activity.

To compare the mean profit share between the two periods, two-sample t-tests (TSTT) were conducted. The TSTT is designed to compare the means of two independent groups. In this study, the groups are not entirely independent, as some companies are present in the top 1,000 list in both periods. This introduces a degree of non-independence, which is a limitation of this analysis.

The TSTT has two main versions. A pooled t-test is conducted if the variances of the two compared groups are similar. However, if the variances differ, Welch's two-sample t-test (WTSTT) is the appropriate choice. As the variance of the two groups usually differs, a one-tailed WTSTT is conducted for each sector. The null hypothesis for this test is that the mean profit share for 2019–2023 is smaller than or equal to the mean profit share for 2008–2012. If the one-tailed P-value is below the standard significance level of 0.05, the null hypothesis is rejected, indicating that the mean profit share in that sector for 2019–2023 is significantly larger than the mean for 2008–2012.

The choice to compare the 2008–2012 and 2019–2023 periods is supported by three arguments. First, the analysis aims to investigate whether there was a rise in rent seeking following the establishment of the new Hungarian political regime, which began when Viktor Orbán's party won a landslide victory in the 2010 elections. Second, comparing individual years could lead to results distorted by macroeconomic cycles. The Hungarian economy experienced a significant downturn between 2009 and 2012, and the early 2020s were also highly volatile. Utilizing five years of data for each period helps to smooth out some of these short-term distortions. Finally, as previously indicated, the sectoral distribution of the top 1,000 firms is unbalanced, meaning some sectors are represented by a limited number of companies. If data from only a single year were used, there would be insufficient observations for certain sectors, potentially rendering the t-test results unreliable.

5. Results

Table 2 presents the WTSTT results for the mean profit share (EBT) of the top 1,000 firms across all sectors. The columns displaying the t-value and the one-tailed p-value are of primary importance for interpreting the results. A positive t-value indicates that the mean profit share was higher in the 2019–2023 period than in 2008–2012, aligning with the hypothesized direction of increase. Conversely, a negative t-value means the opposite (a lower mean in the later period); sectors with negative t-values therefore do not show the hypothesized increase and are not considered to support the research question about an increase. The final column shows the one-tailed p-values. If a p-value is less than the standard significance level of 0.05, the difference between the two means is considered statistically significant at the 5% level.

To visually summarize the findings relevant to the research question (significant increase), cells in the t-value and one-tailed p-value columns are highlighted. A sector's results are highlighted in green if the t-value is positive AND the one-tailed p-value is less than 0.05, indicating a statistically significant increase in mean profit share. Results are highlighted in brown otherwise

(i.e., if the t-value is negative, or if the p-value is greater than 0.05). Based on the WTSTT results and the green highlighting, it can be concluded that the 2019–2023 mean profit share is significantly higher than the 2008–2012 mean in the following sectors: A - Agriculture, Forestry and Fishing; F - Construction; L - Financial and Insurance Activities; M - Real Estate Activities; R - Human Health and Social Work Activities; S - Arts, Sports and Recreation.

Table 2. Welch's two-sample t-test results: mean profit shares for top 1,000 Hungarian firms (2008–2012 vs. 2019–2023)

| Sector | Observations 2008-12 | Mean 2008-12 | Variance 2008-12 | Observations 2019-23 | Mean 2019-23 | Variance 2019-23 | t-value | One-tailed P-value |
|---|---|---|---|---|---|---|---|---|
| A - Agriculture, Forestry and Fishing | 55 | 2.66E-04 | 1.22E-07 | 43 | 1.11E-03 | 1.04E-05 | 1.71 | 0.047 |
| B - Mining and Quarrying | 31 | 4.44E-03 | 6.32E-05 | 17 | 1.44E-03 | 6.46E-06 | -1.93 | 0.030 |
| C - Manufacturing | 1 658 | 1.62E-03 | 1.37E-04 | 1 807 | 1.10E-03 | 2.66E-05 | -1.65 | 0.049 |
| D - Electricity, Gas, Steam and Air Conditioning Supply | 334 | 2.13E-03 | 7.20E-05 | 240 | 1.59E-03 | 5.11E-05 | -0.82 | 0.206 |
| E - Water Supply; Sewerage, Waste Management and Remediation Activities | 63 | 7.11E-04 | 1.29E-06 | 39 | 4.88E-04 | 4.44E-06 | -0.61 | 0.272 |
| F - Construction | 245 | 2.60E-04 | 1.16E-06 | 326 | 1.41E-03 | 2.46E-05 | 4.05 | 0.000 |
| G - Wholesale and Retail Trade; Repair of Motor Vehicles and Motorcycles | 1 443 | 1.32E-04 | 2.50E-05 | 1 568 | 3.92E-04 | 2.17E-05 | 1.47 | 0.071 |
| H - Transportation and Storage | 269 | 5.03E-04 | 1.72E-05 | 268 | 1.62E-04 | 4.39E-05 | -0.72 | 0.237 |
| I - Accommodation and Food Service Activities | 32 | 1.44E-06 | 4.56E-07 | 29 | 1.87E-04 | 9.21E-07 | 0.87 | 0.195 |
| J - Information, Communication | 179 | 1.47E-03 | 7.36E-05 | 201 | 2.65E-03 | 2.55E-04 | 0.91 | 0.183 |
| L - Financial and Insurance Activities | 220 | 9.09E-04 | 3.18E-04 | 121 | 5.94E-03 | 8.16E-05 | 3.45 | 0.000 |
| M - Real Estate Activities | 115 | -1.73E-04 | 1.12E-05 | 22 | 1.16E-03 | 3.62E-06 | 2.61 | 0.006 |
| N - Professional, Scientific and Technical Activities | 176 | 1.42E-03 | 1.09E-05 | 143 | 4.76E-04 | 1.94E-06 | -3.42 | 0.000 |
| O - Administrative and Support Service Activities | 148 | 2.06E-03 | 8.24E-05 | 153 | 8.37E-05 | 2.60E-05 | -2.32 | 0.011 |
| R - Human Health and Social Work Activities | 22 | 4.50E-05 | 5.82E-08 | 5 | 4.79E-04 | 9.01E-08 | 3.02 | 0.015 |
| S - Arts, Sports and Recreation | 10 | 2.35E-03 | 5.60E-06 | 13 | 4.43E-03 | 1.17E-05 | 1.72 | 0.050 |

Source: own calculations

Among the sectors where the mean profit share significantly increased, some have a limited number of observations. Given that the sample covers a five-year period, 5 observations in a sector correspond to just one company, and 20 observations correspond to only four companies. These numbers are considered too low to reliably represent an entire sector, thus disqualifying these sectors from further analysis. With such small sample sizes, detected shifts in profit share may be strongly influenced by individual company performance rather than broader systemic trends. In Table 2, sectors with an insufficient number of observations for the analysis period are highlighted in purple. Consequently, only three sectors remain for further discussion: A - Agriculture, Forestry and Fishing; F - Construction; L - Financial and Insurance Activities.

5.1. Robustness check–comparison with Czech firms

According to the theoretical model presented in Figure 1, the three sectors identified in Hungary experienced a significant rise in their profit share due to artificially created scarcities by political decision-makers. However, this rise in profit shares could have been generated by external factors (such as general macroeconomic trends, shifts in market power of multinational firms, geopolitical developments, etc.) entirely unrelated to rent seeking. The specific method used to check for the effects of rent seeking is the comparison of the mean profit share between two distinct periods. The most effective way to validate this method against the influence of possible external factors is to perform a comparison with a peer economy that is exposed to similar geopolitical and macro-level trends but possesses a different political system that does not lean towards political capitalism.

Hungary's traditional peer economies for comparison are the Visegrád countries. These countries share a similar historical and institutional background, their economic environment is highly comparable, and their level of development is alike. Csaba (2022a) suggests the Visegrád countries are so similar that they can be subsumed under a single model of development. Of the three possible peer economies, Czechia has the indicators that suggest the least amount of regulatory privileges (see Table 3). Therefore, this country is selected for the comparison of sectoral profit shares against the Hungarian results.

Table 3. Selected governance indicators for the Visegrád countries (Scale: 0=worst, 100=best)

| Country | Regulatory quality | | | Rule of law | | |
|---|---|---|---|---|---|---|
|  | 2010 | 2015 | 2020 | 2010 | 2015 | 2020 |
| Czechia | 86.12 | 81.43 | 86.67 | 79.62 | 83.33 | 82.86 |
| Hungary | 80.38 | 74.29 | 66.67 | 72.04 | 64.76 | 66.19 |
| Poland | 81.34 | 79.52 | 76.19 | 68.72 | 76.19 | 67.62 |
| Slovakia | 79.9 | 75.24 | 74.29 | 66.35 | 67.62 | 73.81 |

Source: own compilation based on World Bank Worldwide Government Indicators data

Firm-level data for Czechia are available in a similar structure in CrefoPort as the data used for Hungary. The two datasets share comparable deficiencies: missing NACE codes that must be imputed manually, and an unbalanced sectoral structure where most of the top companies belong either to C - Manufacturing or G - Wholesale and Retail trade. In addition, the Czech data pertaining to the 2020s are incomplete: the threshold net sales value required to be included in the top 1,000 firms drops drastically starting from 2021. For this reason, the second period in which profit shares are calculated only comprises data from three years: 2019, 2020, and 2021. The Czech dataset included some extreme outliers. For instance, the loss of a single company in 2009—Gomanold a.s.—was an order of magnitude larger than the total profits generated by the top 1,000 in that year. Such outliers (fewer than 10 out of the 8,000 data points) were removed. All other elements were calculated in a similar manner to the approach used for Hungary.

The results of the calculations are shown in Table 4. There are six sectors in which the profit share of companies increased between the 2008–12 and 2019–21 periods. This increase is signified by a positive t-value. However, the one-tailed P-value is above the standard 5% significance level in all six cases, which means that there was no sector in Czechia where the profit share of companies, relative to the total profits of the top 1,000 firms, significantly increased.

Table 4. Welch's two-sample t-test results: mean profit shares for top 1,000 Czech firms (2008–2012 vs. 2019–2021)

| Sector | Observations 2008-12 | Mean 2008-12 | Variance 2008-12 | Observations 2019-21 | Mean 2019-21 | Variance 2019-21 | t-value | One-tailed P-value |
|---|---|---|---|---|---|---|---|---|
| A - Agriculture, Forestry and Fishing | 51 | -0.0007 | 1.87E-04 | 55 | 0.0012 | 1.92E-05 | 0.9288 | 0.1784 |
| B - Mining and Quarrying | 51 | 0.0106 | 5.70E-04 | 10 | -0.0001 | 4.14E-08 | -3.2019 | 0.0012 |
| C – Manufacturing | 2044 | 0.0016 | 9.38E-05 | 914 | 0.0002 | 3.88E-05 | -4.445 | 4.58E-06 |
| D - Electricity, Gas, Steam and Air Conditioning Supply | 202 | 0.0089 | 6.24E-04 | 111 | 0.0009 | 9.75E-06 | -4.498 | 5.65E-06 |
| E - Water Supply; Sewerage, Waste Management and Remediation Activities | 84 | 0.0018 | 1.90E-05 | 31 | 0.0002 | 7.34E-08 | -3.3874 | 0.0005 |
| F – Construction | 236 | 0.0018 | 4.48E-05 | 147 | 0.0004 | 2.60E-06 | -3.0822 | 0.0011 |
| G - Wholesale and Retail Trade; Repair of Motor Vehicles and Motorcycles | 1458 | 0.0006 | 3.07E-05 | 769 | 0.0004 | 2.09E-05 | -1.2001 | 0.1151 |
| H - Transportation and Storage | 251 | 0.0004 | 1.15E-04 | 96 | 0.0005 | 4.95E-06 | 0.1787 | 0.4292 |
| I - Accommodation and Food Service Activities | 26 | 1.94E-05 | 7.73E-06 | 53 | 0.0027 | 2.84E-04 | 1.1244 | 0.1327 |
| J - Information, Communication | 140 | 0.0046 | 1.94E-04 | 75 | -0.0004 | 3.48E-05 | -3.6423 | 0.0002 |
| L - Financial and Insurance Activities | 105 | 0.0042 | 1.41E-04 | 54 | 0.0002 | 2.45E-06 | -3.3533 | 0.0005 |
| M - Real Estate Activities | 50 | 0.0028 | 3.77E-04 | 297 | 0.0003 | 4.87E-06 | -0.9024 | 0.1856 |
| N - Professional, Scientific and Technical Activities | 113 | 0.0012 | 9.63E-06 | 135 | 0.0019 | 6.50E-05 | 0.8847 | 0.1887 |
| O - Administrative and Support Service Activities | 87 | 0.0017 | 1.22E-04 | 52 | 5.93E-05 | 4.89E-06 | -1.3751 | 0.0861 |
| R - Human Health and Social Work Activities | 50 | -0.0017 | 4.68E-04 | 96 | 0.0042 | 5.03E-04 | 1.5497 | 0.0621 |
| S - Arts, Sports and Recreation | 33 | 8.99E-05 | 1.03E-04 | 38 | 0.0012 | 3.15E-05 | 0.5639 | 0.2877 |

Source: own calculations

Although Czechia also experienced a democratic backslide in its political system during the latter half of the 2010s, its socio-economic and political environment rendered the country more resistant to political capitalism. While the Czech electorate chose a politician who had accumulated immense economic and media power over several decades (Andrej Babiš), this wealth was amassed prior to his election. Furthermore, he did not develop a bespoke oligarchic network, in contrast to the Hungarian model (Hanley & Vachudova, 2018). The Czech constitutional court has remained relatively free of political capture; the institutional environment makes it difficult to pack the court with loyalists. Consequently, the court has successfully acted as a guardian of political competition, preventing the excessive centralisation of political power (Smekal et al., 2022).

The differences are clear if we concentrate on the sectors where evidence of Hungarian rent seeking was detected in this study. Privatisation, supported by state-provided preferential credit, led to a concentration in land ownership in Hungary. This was possible because the Hungarian state owned 23% of land in 2014 (Gonda, 2019). In contrast, the Land Fund—which managed and privatised land owned by the Czech state—held only 7.5 per cent of the total Czech farmland area in 2010 (Czech News Agency, 2010). Furthermore, although Mr Babiš is the most influential agri-business tycoon in his country, his political career led to a suspension of CAP subsidies paid to his conglomerate, Agrofert (European Parliament, 2021).

In Hungary, one of the main ways to limit competition was through the public procurement system. A study by Bolcha et al. (2024) finds that procurements with less transparency have led to smaller price savings and more frequent single bidders in Czechia. However, the Czech public e-procurement information system enables transparency (Chvalkovská & Skuhrovec, 2010; Nguyen, 2024) and therefore limits the possibility of overpricing. Finally, financial nationalism—another major method of limiting competition and creating scarcities in Hungary—is non-existent in Czechia. Indeed, the Czech banking system remains dominated by foreign-owned banks (OECD, 2022).

When we compare these findings to the Hungarian test results, they suggest that the macroeconomic trends, changes in the market power of multinational firms, and geopolitical developments, which formed the wider economic environment for both Czechia and Hungary over the investigated period, did not cause a major shift in sectoral profit shares. The significant changes detected in Hungary can therefore be attributed to the rent seeking environment of the Hungarian political capitalism.

6. Discussion

Comparable Czech results presented in subsection 5.1 suggest that there are three sectors in Hungary (A - Agriculture, Forestry and Fishing; F - Construction; and L - Financial and Insurance Activities) where rent seeking can be detected. Agriculture is a sector frequently investigated for rent seeking across Europe, with numerous studies addressing the topic. The Hungarian agricultural sector exhibits three key features that raise suspicion of such activities. First, land ownership and land use have become increasingly concentrated. This process began in the 1990s and continued even after 2010. According to Kovách (2018), this concentration largely resulted from political decisions, leading to Hungarian land use concentration being among the highest in the EU by the mid-2010s. Second, in the 2010s, the government sold large quantities of state-owned land, with the majority (up to 70%) acquired by well-connected individuals or groups (Gonda, 2019). Third, large landowners disproportionately received land subsidies (Czyżewski & Matuszczak, 2018). Hungary's share of the Common Agricultural Policy budget amounted to 12.4 billion euros in the 2014–2020 period, one of the highest in the

EU (Gonda, 2019). Consistent with these factors, an increase in profitability is also evident among agricultural firms in the top 1,000 sample. The average return on assets (ROA) ratio for agricultural firms in this sample increased from 4.9% in the 2008–2012 period to 8.4% in the 2019–2023 period (see Table 5).

Table 5. Welch's two-sample t-test results: mean return on assets (ROA) ratios for selected Hungarian firms (2008–2012 vs. 2019–2023)

| Sector | Mean 2008-12 | Variance 2008-12 | Mean 2019-23 | Variance 2019-23 | t-value | One-tailed P-value |
|---|---|---|---|---|---|---|
| A - Agriculture, Forestry and Fishing | 4.9 | 36.2 | 8.4 | 208.9 | 1.48 | 0.072 |
| F – Construction | 3.3 | 141.2 | 10.1 | 366.1 | 5.22 | 0.000 |
| L - Financial and Insurance Activities | 4.5 | 207.6 | 4.2 | 133.3 | -0.21 | 0.418 |
| Total | 2.9 | 459.0 | 6.5 | 565.8 | 7.88 | 0.000 |

Source: own calculations

The construction sector emerged as one of the primary beneficiaries of Hungarian economic policy after 2010. By 2019, the turnover of the broad construction sector reached 42.2 billion euros, marking an 83.6% increase since 2010 (European Commission, 2020). Four key factors strongly suggest a link between the construction sector and rent seeking.

The first factor is the sector's deep connection with the Hungarian public procurement system. In the 2010s, the construction sector accounted for a substantial portion (up to 60-80%) of the total advertised public procurement spending in Hungary (Fazekas et al., 2015). The public procurement system is widely regarded as a key mechanism for channelling public funds to favoured private actors (Szanyi, 2022). Research has shown that the level of competition in government tenders is significantly low, which favours well-connected groups (Tóth & Hajdú, 2018). Furthermore, companies with political connections are more likely to win tenders, and these tenders are typically overpriced (Fazekas et al., 2015).

The second factor is the sector's high profitability in contrast with its low productivity. The total gross operating rate of the broad construction sector was 18.5% in 2018, representing a significant increase from its 2010 value of 12.4%, and higher than the EU-27 average of 16.6% (European Commission, 2020). Consistent with this trend, the mean ROA for firms in the sample almost tripled during the investigated period, rising from 3.3% in the 2008–2012 period to 10.1% in the 2019–2023 period (Table 3). In contrast, the productivity of the construction sector is consistently below the EU average. Moreover, productivity has substantially declined since the early 2000s: the 2023 value was only around 80% of the 2002 value (European Commission, 2023, p. 56).

Factors three and four can be demonstrated using the top 1,000 statistics generated for this study: (3) a complete change in the distribution of major market players, and (4) a drastic drop in construction sector exports. In 2008, the top three construction firms (ranked by profit) were all foreign-owned, and there were only four Hungarian-owned firms in the top ten. By 2023, eight of the top ten were Hungarian-owned. Three of these, including the two most profitable, were owned by the same individual. The only foreign-owned companies remaining in the top ten belong to players involved in projects influenced by foreign sponsors (such as the Samsung battery plant and the Budapest–Belgrade railway developed by Chinese contractors).

The establishment of these new national champions has led to significant market concentration and high profit margins, while simultaneously causing a decline in export activity. If heavily overpriced domestic contracts can be secured through public procurement without competition, there is little incentive to pursue overseas projects. In 2008, the construction firms included in the top 1,000 accounted for a 0.28 per cent share of Hungary's total exports. By 2022, this share had fallen more than tenfold to a mere 0.02 per cent.

Financial nationalism constituted one of the focal points of the new regime's economic policy strategy; the Prime Minister openly communicated the government's aim to increase domestic ownership in the banking sector to over 50% (Sebők & Simons, 2022). This aim was accomplished, and the manner in which it was achieved suggests the presence of rent seeking. The reshaping of the bank structure involved four key steps of government intervention and favouritism. First, the government exerted pressure on some smaller financial cooperatives and larger foreign-owned banks by changing regulations and introducing sectoral taxes (Oellerich, 2022; Szanyi, 2022). Second, a process of re-nationalization was initiated, during which the government intervened in the ownership structure of seven banks, including four top banks with the highest balance sheets (Sebők & Simons, 2022). Third, major banks were then re-privatized, often to entities identified as 'national capitalists'. Fourth, these re-privatizations were typically funded with credits provided by state-controlled financial institutions (Sebők & Simons, 2022). These actions resulted in a more concentrated market structure within the banking sector. While higher market power typically generates higher profits, the overall efficiency of the Hungarian banking sector reportedly decreased during the 2010s (Székely, 2018).

Table 6 summarizes evidence from the literature that supports this paper's argument for excessive rent seeking in three Hungarian sectors: agriculture, construction, and finance.

Table 6. Signs of rent seeking in the three identified Hungarian sectors

| Sector | Own findings | Backing literature |
|---|---|---|
| A - Agriculture, Forestry and Fishing | Mean profit (EBT) share increased by 320% (p-value=0.04) Mean ROA increased by 70% (p-value=0.07) | Land ownership and land use concentration (Kovách, 2018) Majority of state-owned land sold to well-connected groups (Gonda, 2019) Disproportionately high receipts of land subsidies by large owners (Czyżewski & Matuszczak, 2018) |
| F – Construction | Mean profit (EBT) share increased by 440% (p-value=0.00) Mean ROA increased by 207% (p-value=0.00) | Sector received more than half of public procurement spending (Fazekas et al., 2015) Public procurement favours well-connected groups, low competition, common overpricing (Fazekas et al., 2015; Szanyi, 2022; Tóth & Hajdú, 2018) Contrast between high profitability and low productivity (European Commission, 2023) |
| L - Financial and Insurance Activities | Mean profit (EBT) share increased by 553% (p-value=0.00) Mean ROA decreased | Government aim of 'financial nationalism' (Sebők & Simons, 2022) Regulatory pressure on bank owners (Oellerich, 2022; Szanyi, 2022) |

| | | Re-nationalisation, and then re-privatisation with state credit (Sebők & Simons, 2022) High profitability and low efficiency (Székely, 2018) |

Source: own work

This study points to suggestive evidence of how the Hungarian system of political capitalism encourages rent seeking, thereby hindering long-term growth. Six linked (Congleton, 2024) mechanisms are discussed:

1. Excessive concentration of political power, which enables the implementation of mechanisms that limit transparency. This facilitates the provision of regulatory privileges to selected groups and the formation of political-business networks without significant political costs. While institutional restraints slowed this process in Czechia, the concentration of political power in Hungary is highly apparent.
2. Regulatory privileges and political-business networks restrict competition, leading to higher market concentration and increased market power. The most prominent manifestation of this limited competition was the emergence of national champions during the 2010s.
3. In agriculture, vast tracts of state-owned land were privatised through tenders that favoured influential groups. These groups often received state-provided credit to facilitate land purchases, leading to a concentration of both market power and CAP subsidy receipts.
4. In construction, Hungarian-owned national champions emerged, aided by a public procurement system that restricted the number of bidders through narrowly defined requirements. Single-bidder contracts became frequent, and the resulting lack of competition led to significant overpricing.
5. In the financial sector, regulatory changes increased the pressure on owners to sell assets. Nationalisation followed by reprivatisation was employed to establish national champions, with favoured groups often utilising state-provided credit for these acquisitions. This reshuffling of ownership further concentrated market power.
6. Finally, the increase in market power allowed dominant firms to raise prices, a process that led to increased profits in the agricultural, construction, and financial sectors. This increased profit share is identified in this paper as empirical evidence of rent seeking.

7. Conclusion

The purpose of this paper was to evaluate political capitalism in Hungary by examining the impact of political decisions on rents. The study compared the mean profit share (earnings before tax) of the top 1,000 Hungarian firms across two periods: 2008–2012 and 2019–2023. The top 1,000 firms were selected each year based on net sales value. While the specific firms comprising the top 1,000 changed over time, a substantial number (445 firms) were present on both the 2008 and 2023 lists. These firms collectively account for a significant share of Hungarian economic output, employing 21% of the total labour force in 2023.

The analysis assumed that a significant increase in the mean profit share for firms within a specific sector could be interpreted as evidence of increased rent seeking in that sector. The theoretical model used for the analysis suggests that regulatory privileges and the political-business networks of the Hungarian political capitalism create scarcities in the markets of certain sectors. These sectoral scarcities drive prices up and ultimately increase the profits of firms operating in those sectors. Applying Welch's two-sample t-tests, six sectors initially showed a statistically significant increase in mean profit share. However, three of these were excluded from further consideration due to an insufficient number of observations in the sample, as the financial data of a single company could disproportionately influence the results. Following this exclusion, three sectors were identified as potentially experiencing increased rent seeking during the investigated period: 1) agriculture, forestry and fishing, 2) construction, and 3) financial and insurance activities.

In the agricultural sector, the mean profit share increased by 320%, while the mean return on assets increased by 70% in the sample between the two periods. The construction sector experienced a period of rapid expansion after 2010. The number of construction firms included in the sample in these sectors increased from 245 (total for 2008–2012) to 326 (total for 2019–2023). The mean profit share for these firms increased more than fivefold, and the mean ROA increased from 3.3% in 2008–2012 to 10.1% in 2019–2023. The financial sector also experienced a phase of concentration, with the total number of observed firms per period decreasing from 220 to 121; however, the mean profit share for firms in this sector increased by more than 6.5 times.

To check whether the increases in the profit shares of the three sectors could have been caused by external factors, a similar analysis was conducted using Czech data. The results show that there were no statistically significant increases in the profit share of any Czech sector during the investigated period, which substantially strengthens the primary findings of this study. A review of the relevant literature provides further support for these findings by highlighting specific mechanisms of state intervention and their impact on these sectors. Evidence from other studies indicates an increased amount of state intervention, including changes in regulation to initiate ownership restructuring, re-nationalisation followed by re-privatisation to favoured groups using state-provided credit, concentration of market power through government actions, decreased competition in tenders via narrowly defined requests for proposals, and overpricing in the public procurement system. The literature also documents signs of low efficiency in the construction and financial sectors despite rising profits.

This study is subject to three main limitations. First, the analysis assumes that observed increases in company profit share can be interpreted as evidence of rent seeking and subsequent rent capture. Theoretically, in perfectly competitive markets, potential rents should be fully dissipated through the costs of rent seeking, leaving no excess profit. However, in real-world settings, and particularly under political capitalism where political decisions may protect rent seekers, rent dissipation is incomplete. The comparison with Czechia provided evidence that the rise in profit share is likely not to have been caused by external factors impacting all countries in the region. However, there could be possible factors that are unrelated to the Hungarian political capitalism regime but are country-specific (e.g., productivity and management issues, industrial organisation). The method utilised does not allow us to ascertain the impact of these specific country-level factors.

Second, the unbalanced distribution of firms within the top 1,000 sample resulted in some sectors having a very limited number of observations. Although the t-test indicated a significant increase in mean profit share for three sectors, the small sample size means that the inference of increased rent seeking in those sectors could not be considered conclusive.

Third, the statistical analysis utilized a two-sample t-test, which assumes that the two samples being compared are independent. This assumption was violated in this study, as a substantial number of companies were included in the sample in both the 2008–2012 and 2019–2023 periods.

To address the sample size limitation, future studies could increase the overall sample size beyond the top 1,000 firms. Addressing the violation of the independence assumption would require employing statistical methods designed for panel data.

Data sources

CrefoPort: https://www.crefoport.hu/

Hungarian Central Statistics Office: https://www.ksh.hu/stadat?lang=hu&theme=gdp

World Bank Worldwide Governance Indicators: https://www.worldbank.org/en/publication/worldwide-governance-indicators/interactive-data-access